# Preformed Cooper pairs in layered FeSe-based superconductors

B. L. Kang[1,†], M. Z. Shi[1,†], S. J. Li[1], H. H. Wang[1], Q. Zhang[1], D. Zhao[1], J. Li[1], D. W. Song[1], L. X. Zheng[1], L. P. Nie[1], T. Wu[1,2,3,4,*] and X. H. Chen[1,2,3,4,*]

1. Hefei National Laboratory for Physical Sciences at the Microscale and Department of Physics, and Key Laboratory of Strongly-coupled Quantum Matter Physics, Chinese Academy of Sciences, University of Science and Technology of China, Hefei, Anhui 230026, China

2. CAS Center for Excellence in Superconducting Electronics (CENSE), Shanghai 200050, China

3. CAS Center for Excellence in Quantum Information and Quantum Physics, Hefei, Anhui 230026, China

4. Collaborative Innovation Center of Advanced Microstructures, Nanjing University, Nanjing 210093, China

† These authors contributed equally to this work

*Corresponding author. Email: wutao@ustc.edu.cn ; chenxh@ustc.edu.cn

**Superconductivity arises from two distinct quantum phenomena: electron pairing and long-range phase coherence. In conventional superconductors, the two quantum phenomena generally take place simultaneously, while the electron pairing occurs at higher temperature than the long-range phase coherence in the underdoped high-$T_c$ cuprate superconductors. Recently, whether electron pairing is also prior to long-range phase coherence in single-layer FeSe film on SrTiO$_3$ substrate is under debate. Here, by measuring Knight shift and nuclear spin-lattice relaxation rate, we unambiguously reveal a pseudogap behavior below $T_p$ ~ 60 K in two layered FeSe-based superconductors with quasi-two-dimension. In the pseudogap regime, a weak diamagnetic signal and a remarkable Nernst effect are also observed, which indicate that the observed pseudogap behavior is related to superconducting fluctuations. These works confirm that strong phase fluctuation**

**is an important character in the two-dimensional iron-based superconductors as widely observed in high-$T_c$ cuprate superconductors.**

Phase fluctuation is an important character for high-$T_c$ superconductivity (SC)(*1*), and determines the condensation of Cooper pairs. In conventional Bardeen-Cooper-Schrieffer (BCS) superconductors, the superfluid density is so large that the pairing and condensation of Cooper pairs always happen simultaneously, and the superconducting transition temperature ($T_c$) is determined by the pairing temperature ($T_p$). However, in the underdoped high-$T_c$ cuprate superconductors, which were treated as a doped Mott insulator(*2*), the famous "Uemura relation" indicates a close correlation between the superfluid density and the $T_c$(*3*), suggesting an important role of phase fluctuation on the determination of $T_c$(*1*). Previous studies also observed significant superconducting fluctuations above $T_c$, supporting a preformed pairing scenario(*4-7*). In contrast, the signature of preformed pairing in another high-$T_c$ family, iron-based superconductors, is still elusive(*8-13*). The discovery of large pairing gap in the single-layer FeSe film on SrTiO$_3$ substrate sheds light on this issue(*14*). The interface as a possible origin to enhance superconducting pairing in the single-layer FeSe film on SrTiO$_3$ has been widely discussed(*15-21*). In addition, the angle-resolved photoemission spectroscopy(ARPES) experiments revealed that the closing temperature of the pairing gap in single-layer FeSe film is up to 65 K(*22,23*), while the zero-resistance transition ($T_{c0}$) determined by transport measurement is observed only below 40 K(*24-27*). A natural explanation on such phenomenon is related to the preformed pairing analogous to that in the underdoped cuprate superconductors.

Very recently, organic molecule intercalated FeSe superconductors (CTA)$_x$FeSe(*28*) and (TBA)$_x$FeSe(*29*) with $T_{c0}$ ~ 43 K have been synthesized by us. As shown in Fig.1A, the charge from organic ion tetrabutyl ammonium (TBA$^+$) is transferred to FeSe layers, which is important to achieve the high-$T_c$ superconductivity in FeSe layer(*17,22,25,30*). The distance between adjacent FeSe layers is enlarged from 5.5 Å in the pristine FeSe to 15.5 Å in (TBA)$_x$FeSe due to the intercalation of TBA molecules. The intercalation of organic molecule makes the (TBA)$_x$FeSe superconductor two-dimensional (2D). In these 2D FeSe-based superconductors, we unambiguously reveal an intrinsic pseudogap behavior below $T_p \sim 60$ K by measuring Knight shift and nuclear spin-lattice relaxation rate. In addition, a weak 2D diamagnetic signal and a remarkable Nernst effect are also observed in the pseudogap regime. These results definitely verify the preformed Cooper pairs in the 2D FeSe-based superconductors. It should be the same scenario for the single-layer FeSe film.

As we know, nuclear magnetic resonance (NMR) is a local probe to measure spin susceptibility ($\chi_s$). When the electrons are bound into Cooper pairs with a spin-singlet pairing, the local spin susceptibility will diminish below pairing temperature. Even if the long-range coherence among Cooper pairs is lost due to strong phase fluctuations, a drop in the local spin susceptibility is still expected(*31*). Therefore, NMR is a quite sensitive probe for superconducting pairing. In NMR measurement, both of the Knight shift (*K*) and the nuclear spin-lattice relaxation rate ($1/T_1$) are related to the spin susceptibility. Usually, the Knight shift is related to the uniform spin susceptibility with $K_{tot} = A\chi_s + K_{orb}$, where $K_{orb}$ is the orbital contribution and is always temperature independent, and A is the hyperfine coupling tensor between nuclear and electronic spins. The nuclear spin-lattice relaxation rate ($1/T_1$) is related to dynamic spin susceptibility with $\frac{1}{T_1 T} \sim \sum_q \gamma_n^2 |A_\perp(q)|^2 \frac{\chi''(q,\omega)}{\omega}$, where $\chi''(q,\omega)$ is the *q*-dependent imaginary of dynamic spin susceptibility and $A_\perp(q)$ is the *q*-dependence of hyperfine form factor (more explanation see the Methods). Usually, $\frac{1}{T_1 T} = \frac{1}{T_1^{QP} T} + \frac{1}{T_1^{SF} T}$, the first term is from quasiparticles and the second term is from additional spin fluctuations. In a Fermi liquid picture, which ignores the *q*-dependent spin fluctuations, both of the quantities keep a simple Korringa relation with $\frac{1}{T_1 T} \sim K^2 \sim N(E_F)^2$(*32*), where *N(E_F)* is the density of state at Fermi level. In this sense, both Knight shift and nuclear spin-lattice relaxation rate are practical to identify pseudogap phenomenon(*33*).

As shown in Fig.2A, the temperature-dependent Knight shift of $^{77}$Se nuclei shows a clear deviation from the high-temperature behavior below 60 K, which is more evident in the differential curve. If we consider that the superconductivity is the only active electronic instability at low temperature, such deviation must be related to superconducting pairing. As shown in Fig.2C, a clear shift of NMR spectrum to low frequency without any change of profile indicates an intrinsic suppression of the uniform spin susceptibility below 60 K. A similar deviation from high-temperature behavior is also observed in the temperature-dependent $1/T_1T$ (Fig.2B), which is in agreement with the Knight shift results. Moreover, the temperature-dependent stretching exponent of $T_1$ fitting shown in Fig.2D indicates a homogeneous spin dynamics in NMR time scale above $T_{c0}$. Therefore, the present NMR results definitely support an intrinsic pseudogap behavior with the characteristic temperature $T_p \sim 60$ K. It should be emphasized that the similar temperature-dependent behavior between Knight shift and $1/T_1T$ also confirms that the observed pseudogap behavior is from quasiparticles instead of spin fluctuations. As shown in Fig.2E, the conductance also exhibits a clear enhancement below 60 K, which is consistent with a fluctuating superconductivity above $T_{c0}$. In addition, it should be pointed out that the observed pseudogap behavior in NMR measurement is independent on field orientation and field strength(see details in Supplementary Materials S2). These results are consistent with a preformed pairing picture with considerable pairing gap suggested by scanning tunneling microscope (STM) measurement shown in Fig.1D.

On the other hand, whether a pseudogap behavior in NMR measurement can be unambiguously ascribed to the preformed pairing is not straightforward if there are other electronic instabilities, such as charge/spin density wave. Here, in order to further clarify the nature of pseudogap behavior below $T_p$, the bulk magnetization was measured with external magnetic field applied in-plane and out-of-plane, respectively. Usually, the weak diamagnetic signal due to the preformed pairing can be detected by the bulk magnetization measurement(*34*). As shown in Fig.2F, the high-field magnetization shows a weak diamagnetic signal well above $T_{c0}$ with magnetic field applied along c-axis. The onset temperature of the weak diamagnetic signal is around $T_p \sim 60$ K, perfectly consistent with the NMR results. It strongly confirms that a 2D superconducting fluctuation emerges below $T_p$, being similar to that in the single-layer FeSe film(*35*). It is well known that the Nernst effect is another quite sensitive probe for superconducting fluctuations(*5*). As shown in Fig.2G, a remarkable Nernst signal is observed above $T_{c0}$ under an external magnetic field of 13.5 T applied along c-axis. At higher temperature, the Nernst signal is very small and almost temperature-independent. Below about 65 K, the Nernst signal shows a clear increase and then reaches a maximum around 40 K. Below 40 K, the Nernst signal continuously decreases. Such temperature dependence of Nernst effect was widely observed in the underdoped cuprate superconductors, and the Nernst signal well above $T_{c0}$ is usually ascribed to the free vortex contribution(common Gaussian fluctuation only appears around $T_{c0}$)(*5*). The Nernst effect further supports a persistent superconducting fluctuation above $T_{c0}$, consistent with the results of NMR and diamagnetism.

As shown in Fig.3A and B, we also measured the anisotropy of resistivity between c-axis and ab-plane. Compared to bulk FeSe, the value of anisotropy is enhanced by about five orders of magnitude, suggesting an intercalation-induced dimensional crossover from 3D to 2D. Considering such highly 2D electronic structure, a natural explanation for the observed pseudogap behavior in these layered FeSe-based superconductors is due to strong phase fluctuation in 2D limit(*24*), in which the spontaneous topological excitations (vortex) due to strong phase fluctuation destroy the zero-resistance state below the 2D superconducting pairing temperature ($T_c^{2D}$). Based on the Berezinskii-Kosterlitz-Thouless (BKT) theory, the zero-resistance state only emerges when vortex and antivortex are bound into pairs below a so-called BKT transition temperature ($T_{BKT}$)(*36,37*). Such topological transition due to the unbinding of vortex-antivortex pairs manifests a jump of the power-law exponent in the current-voltage (*I-V*) characteristic curves and a disappearance of Ohmic resistance obeying the Halperin–Nelson scaling law(*38*), which are important manifestations of 2D superconductivity(*29*). To further verify the nature of both zero-resistance transition and pseudogap phenomenon, we also measured the temperature-dependent *I-V* curves across the $T_{c0}$. As shown in Fig.3C, a power-law transition with $V \sim I^\alpha$ can be identified during the zero-resistance transition. We extracted the temperature dependence of the power-law exponent $\alpha$, which was deduced by fitting the *I-V* curves, as shown in Fig. 3D. At $T = 44$ K, the exponent $\alpha$ continuously approaches to the value of 3, which can be used to define a supposed BKT transition(*39*). Moreover, the temperature-dependent resistance R(*T*) follows a typical BKT-like behavior with $R(T) = R_0\exp[-b/(T-T_{BKT})^{1/2}]$ in the temperature range close to $T_{BKT}$, where $R_0$ and b are material dependent parameters(*38*). As shown in Fig. 4E, the extracted value of $T_{BKT}$ from the measured R(*T*) curve is about 44.7 K, in agreement with the *I-V* results. Both anisotropic transport and *I-V* curves support a 2D-like (or BKT-like) nature in these layered FeSe-based superconductors, which hints an important role of phase fluctuation.

Up to now, all experimental results by different techniques support a pseudogap behavior due to the preformed pairing below $T_p \sim 60$ K and a highly anisotropic superconductivity below $T_{c0} \sim 43$ K in (TBA)$_x$FeSe. A summarized *H-T* phase diagram is shown in Fig. 4. The similar pseudogap behavior is also observed in another layered FeSe-based superconductor (CTA)$_x$FeSe (more details see Supplementary Materials S3). These results confirm that such pseudogap behavior is intrinsic for the layered FeSe-based superconductors. In addition, STM measurement indicates that the local superconducting gap is about 16 meV in these layered FeSe-based superconductors (see Fig.1D for (TBA)$_x$FeSe and Fig.S5 for (CTA)$_x$FeSe in Supplementary Materials), which are comparable to that of the single-layer FeSe film with the gap value varying from 13 to 20 meV (see Fig.S8 in the Supplementary Materials S4). Therefore, the much higher pairing temperature ($\sim 65$ K) than zero-resistance temperature ($< 40$ K) in the single-layer FeSe film might be also ascribed to the same preformed pairing scenario. These works indicate that strong phase fluctuation is an important character in 2D iron-based superconductors as widely observed in high-$T_c$ cuprate superconductors. How to understand the underlying physics behind preformed pairing is relevant to high-$T_c$ mechanism and brings a challenge to theory. Finally, our work not only reveals preformed Cooper pairs, but also suggests a possible enhancement of pairing strength by dimensional crossover from 3D to 2D.

**Data availability**

The data that support the plots within this paper and other findings of this study are available from the corresponding authors upon reasonable request.

**Reference**

1. V. J. Emery, S. A. Kivelson, Importance of Phase Fluctuations in Superconductors with Small Superfluid Density. *Nature* **374**, 434-437 (1995).
2. P. A. Lee, N. Nagaosa, X.G. Wen, Doping a Mott insulator: Physics of high-temperature superconductivity. *Rev. Mod. Phys.* **78**, 17-85 (2006).
3. Y. J. Uemura, G. M. Luke, B. J. Sternlieb, J. H. Brewer, J. F. Carolan, W. N. Hardy, R. Kadono, J. R. Kempton, R. F. Kiefl, S. R. Kreitzman, P. Mulhern, T. M. Riseman, D. L. Williams, B.X. Yang, S. Uchida, H. Takagi, J. Gopalakrishnan, A. W. Sleight, M. A. Subramanian, C. L. Chien, M. Z. Cieplak, G. Xiao, V. Y. Lee, B. W. Statt, C. E. Stronach, W. J. Kossler, X.H. Yu, Universal Correlations between $T_c$ and $n_s/m^*$ (Carrier Density over Effective Mass) in High-$T_c$ Cuprate Superconductors. *Phys. Rev. Lett.* **62**, 2317-2320 (1989).
4. L. Li, Y.Y. Wang, S. Komiya, S. Ono, Y. Ando, G.D. Gu, N. P. Ong, Diamagnetism and Cooper pairing above $T_c$ in cuprates. *Phys. Rev. B* **81**, 054510 (2010).
5. Y.Y. Wang, L. Li, N. P. Ong, Nernst effect in high-$T_c$ superconductors. *Phys. Rev. B* **73**, 024510 (2006).
6. J. L. Tallon, J. G. Storey, J. W. Loram, Fluctuations and critical temperature reduction in cuprate superconductors. *Phys. Rev. B* **83**, 092502 (2011).
7. J. Corson, R. Mallozzi, J. Orenstein, J. N. Eckstein, I. Bozovic, Vanishing of phase coherence in underdoped $Bi_2Sr_2CaCu_2O_{8+\delta}$. *Nature* **398**, 221-223 (1999).
8. S. Kasahara, T. Yamashita, A. Shi, R. Kobayashi, Y. Shimoyama, T. Watashige, K. Ishida, T. Terashima, T. Wolf, F. Hardy, C. Meingast, H. Von Lohneysen, A. Levchenko, T. Shibauchi, Y. Matsuda, Giant superconducting fluctuations in the compensated semimetal FeSe at the BCS-BEC crossover. *Nat.Commun.* **7**, 12843 (2016).

9. H. Takahashi, F. Nabeshima, R. Ogawa, E. Ohmichi, H. Ohta, A. Maeda, Superconducting fluctuations in FeSe investigated by precise torque magnetometry. *Phys. Rev. B* **99**, 060503 (2019).

10. K. Ahilan, F. L. Ning, T. Imai, A. S. Sefat, R. Jin, M. A. McGuire, B. C. Sales, D. Mandrus, $^{19}$F NMR investigation of the iron pnictide superconductor LaFeAsO$_{0.89}$F$_{0.11}$. *Phys. Rev. B* **78**, 100501 (2008).

11. T. Sato, S. Souma, K. Nakayama, K. Terashima, K. Sugawara, T. Takahashi, Y. Kamihara, M. Hirano, H. Hosono, Superconducting gap and pseudogap in iron-based layered superconductor La(O$_{1-x}$F$_x$)FeAs. *J. Phys. Soc. Jpn.* **77**, 063708 (2008).

12. Z.W. Zhu, Z.A. Xu, X. Lin, G.H. Cao, C.M. Feng, G.F. Chen, Z. Li, J.L. Luo, N.L. Wang, Nernst effect of a new iron-based superconductor LaO$_{1-x}$F$_x$FeAs. *New J. Phys.* **10**, 063021 (2008).

13. A. L. Shi, T. Arai, S. Kitagawa, T. Yamanaka, K. Ishida, A. E. Bohmer, C. Meingast, T. Wolf, M. Hirata, T. Sasaki, Pseudogap Behavior of the Nuclear Spin-Lattice Relaxation Rate in FeSe Probed by $^{77}$Se NMR. *J. Phys. Soc. Jpn.* **87**, 013704 (2018).

14. Q.Y. Wang, Z. Li, W.H. Zhang, Z.C. Zhang, J.S. Zhang, W. Li, H. Ding, Y.B. Ou, P. Deng, K. Chang, J. Wen, C.L. Song, K. He, J.F. Jia, S.H. Ji, Y.Y. Wang, L.L. Wang, X. Chen, X.C. Ma, Q.K. Xue, Interface-Induced High-Temperature Superconductivity in Single Unit-Cell FeSe Films on SrTiO$_3$. *Chin. Phys. Lett.* **29**, 037402 (2012).

15. J. J. Lee, F. T. Schmitt, R. G. Moore, S. Johnston, Y.T. Cui, W. Li, M. Yi, Z.K. Liu, M. Hashimoto, Y. Zhang, D.H. Lu, T. P. Devereaux, D.H. Lee, Z.X. Shen, Interfacial mode coupling as the origin of the enhancement of $T_c$ in FeSe films on SrTiO$_3$. *Nature* **515**, 245-248 (2014).

16. D. H. Lee, What makes the $T_c$ of FeSe/SrTiO$_3$ so high? *Chin. Phys. B* **24**, 117405 (2015).

17. Y. Miyata, K. Nakayama, K. Sugawara, T. Sato, T. Takahashi, High-temperature superconductivity in potassium-coated multilayer FeSe thin films. *Nat. Mater.* **14**, 775-779 (2015).

18. L. Rademaker, Y. Wang, T. Berlijn, S. Johnston, Enhanced superconductivity due to

forward scattering in FeSe thin films on SrTiO₃ substrates. *New J. Phys.* **18**, 022001 (2016).

19. Y. J. Zhou, A. J. Millis, Dipolar phonons and electronic screening in monolayer FeSe on SrTiO₃. *Phys. Rev. B* **96**, 054516 (2017).

20. Q. Song, T. L. Yu, X. Lou, B. P. Xie, H. C. Xu, C. H. P. Wen, Q. Yao, S. Y. Zhang, X. T. Zhu, J. D. Guo, R. Peng, D. L. Feng, Evidence of cooperative effect on the enhanced superconducting transition temperature at the FeSe/SrTiO₃ interface. *Nat. Commun.* **10**, 758 (2019).

21. F. M. Li, G. A. Sawatzky, Electron Phonon Coupling versus Photoelectron Energy Loss at the Origin of Replica Bands in Photoemission of FeSe on SrTiO₃. *Phys. Rev. Lett.* **120**, 237001 (2018).

22. S.L. He, J.F. He, W.H. Zhang, L. Zhao, D.F. Liu, X. Liu, D.X. Mou, Y.B. Ou, Q.Y. Wang, Z. Li, L.L. Wang, Y.Y. Peng, Y. Liu, C. Y. Chen, L. Yu, G.D. Liu, X.L. Dong, J. Zhang, C.T. Chen, Z.Y. Xu, X. Chen, X. Ma, Q.K. Xue, X.J. Zhou, Phase diagram and electronic indication of high-temperature superconductivity at 65 K in single-layer FeSe films. *Nat. Mater.* **12**, 605-610 (2013).

23. S.Y. Tan, Y. Zhang, M. Xia, Z.R. Ye, F. Chen, X. Xie, R. Peng, D.F. Xu, Q. Fan, H.C. Xu, J. Jiang, T. Zhang, X.C. Lai, T. Xiang, J.P. Hu, B.P. Xie, D.L. Feng, Interface-induced superconductivity and strain-dependent spin density waves in FeSe/SrTiO₃ thin films. *Nat. Mater.* **12**, 634-640 (2013).

24. W.H. Zhang, Y. Sun, J.S. Zhang, F.S. Li, M.H. Guo, Y.F. Zhao, H.M. Zhang, J.P. Peng, Y. Xing, H.C. Wang, T. Fujita, A. Hirata, Z. Li, H. Ding, C.J. Tang, M. Wang, Q.Y. Wang, K. He, S.H. Ji, X. Chen, J.F. Wang, Z.C. Xia, L. Li, Y.Y. Wang, J. Wang, L.L. Wang, M.W. Chen, Q.K. Xue, X.C. Ma, Direct Observation of high-temperature superconductivity in one-unit-cell FeSe films. *Chin. Phys. Lett.* **31**, 017401 (2014).

25. W.H. Zhang, Z. Li, F.S. Li, H.M. Zhang, J.P. Peng, C.J. Tang, Q.Y. Wang, K. He, X. Chen, L.L. Wang, X.C. Ma, Q.K. Xue, Interface charge doping effects on superconductivity of single-unit-cell FeSe films on SrTiO₃ substrates. *Phys. Rev. B* **89**, 060506 (2014).

26. Y. Sun, W.H. Zhang, Y. Xing, F.S. Li, Y.F. Zhao, Z.C. Xia, L.L. Wang, X.C. Ma,

Q.K. Xue, J. Wang, High temperature superconducting FeSe films on SrTiO$_3$ substrates. *Sci. Rep.* **4**, 6040 (2014).

27. Q.Y. Wang, W.H. Zhang, Z.C. Zhang, Y. Sun, Y. Xing, Y.Y. Wang, L.L. Wang, X.C. Ma, Q.K. Xue, J. Wang, Thickness dependence of superconductivity and superconductor-insulator transition in ultrathin FeSe films on SrTiO$_3$(001) substrate. *2D Mater.* **2**, 044012 (2015).

28. M. Z. Shi, N. Z. Wang, B. Lei, C. Shang, F. B. Meng, L. K. Ma, F. X. Zhang, D. Z. Kuang, X. H. Chen, Organic-ion-intercalated FeSe-based superconductors. *Phys. Rev. Mater.* **2**, 074801 (2018).

29. M. Z. Shi, N. Z. Wang, B. Lei, J. J. Ying, C. S. Zhu, Z. L. Sun, J. H. Cui, F. B. Meng, C. Shang, L. K. Ma, X. H. Chen, FeSe-based superconductors with a superconducting transition temperature of 50 K. *New J. Phys.* **20**, 123007 (2018).

30. B. Lei, J. H. Cui, Z. J. Xiang, C. Shang, N. Z. Wang, G. J. Ye, X. G. Luo, T. Wu, Z. Sun, X. H. Chen, Evolution of High-Temperature Superconductivity from a Low-T$_c$ Phase Tuned by Carrier Concentration in FeSe Thin Flakes. *Phys. Rev. Lett.* **116**, 077002 (2016).

31. C. Berthier, M. H. Julien, M. Horvatic, Y. Berthier, NMR studies of the normal state of high temperature superconductors. *J. Phys. I France* **6**, 2205-2236 (1996).

32. J. Korringa, Nuclear Magnetic Relaxation and Resonance Line Shift in Metals. *Physica* **16**, 601-610 (1950).

33. H. Alloul, T. Ohno, P. Mendels, $^{89}$Y NMR Evidence for a Fermi-Liquid Behavior in YBa$_2$Cu$_3$O$_{6+x}$. *Phys. Rev. Lett.* **63**, 1700-1703 (1989).

34. Q. Li, M. Hucker, G.D. Gu, A. M. Tsvelik, J. M. Tranquada, Two-dimensional superconducting fluctuations in stripe-ordered La$_{1.875}$Ba$_{0.125}$CuO$_4$. *Phys. Rev. Lett.* **99**, 067001 (2007).

35. Z. C. Zhang, Y. H. Wang, Q. Song, C. Liu, R. Peng, K. A. Moler, D.L. Feng, Y.Y. Wang, Onset of the Meissner effect at 65 K in FeSe thin film grown on Nb-doped SrTiO$_3$ substrate. *Sci. Bull.* **60**, 1301-1304 (2015).

36. V. L. Berezinskii, Destruction of long-range order in one-dimensional and two-dimensional systems having a continuous symmetry group I. Classical Systems. *Sov.*

*Phys.JETP* **32**, 493-500 (1971). II. Quantum Systems. *Sov. Phys. JETP* **34**, 610-617 (1972).

37. J. M. Kosterlitz, D. J. Thouless, Ordering, metastability and phase-transitions in two dimensional systems. *J. Phys. C Solid State* **6**, 1181-1203 (1973).

38. B. I. Halperin, D. R. Nelson, Resistive transition in superconducting films. *J. Low Temp. Phys.* **36**, 599-616 (1979).

39. N. Reyren, S. Thiel, A. D. Caviglia, L. F. Kourkoutis, G. Hammerl, C. Richter, C. W. Schneider, T. Kopp, A. S. Ruetschi, D. Jaccard, M. Gabay, D. A. Muller, J. M. Triscone, J. Mannhart, Superconducting interfaces between insulating oxides. *Science* **317**, 1196-1199 (2007).

40. S. I. Vedeneev, B. A. Piot, D. K. Maude, A. V. Sadakov, Temperature dependence of the upper critical field of FeSe single crystals. *Phys. Rev. B* **87**, 134512 (2013).

41. G. Blatter, M. V. Feigelman, V. B. Geshkenbein, A. I. Larkin, V. M. Vinokur, Vortices in high-temperature superconductors. *Rev. Mod. Phys.* **66**, 1125-1388 (1994).

42. B. L. Kang et al., Replication data for: Preformed-pairing-induced pseudogap in layered FeSe-based superconductors, version 1.0, xxxxx (2019).

**Acknowledgments**

We thank Z. J. Xiang for valuable suggestionon Nernst measurement, F. C. Zhang and Z. Y. Weng for insightful discussion. This work is supported by the National Natural Science Foundation of China (Grants No. 11888101, 11522434), the National Key R&D Program of the MOST of China (Grants No. 2016YFA0300201, 2017YFA0303000), the Strategic Priority Research Program of Chinese Academy of Sciences (Grant No. XDB25000000), the Anhui Initiative in Quantum Information Technologies (Grant No. AHY160000).

**Author contributions**

T.W. and X.H.C conceived and designed the experiments. B.L.K. and S.J.L. performed NMR measurements with the assistance from T.W., D.Z., J.L., D.W.S., L.X.Z. and

L.P.N., M.Z.S. and B.L.K. performed sample growth, transport measurement and bulk magnetization measurement. Q.Z. performed STM measurement. H.H.W. performed Nernst measurement. T.W. and X.H.C. interpret the data. T.W. and X.H.C. wrote the manuscript. All authors discussed the results and commented on the manuscript.

## Methods

**Crystal growth and characterization.** The synthesis of $(TBA)_x FeSe$ includes two steps. Firstly, the shining plate-like single crystals of tetragonal FeSe were grown using a $KCl-AlCl_3$ flux technique as described in detail elsewhere*(43)*. Secondly, the $(TBA)_x FeSe$ single crystals were synthesized through an electrochemical intercalation process using FeSe as the starting material. Since the grown crystals are very sensitive to oxygen and moisture, all the processes including sample growth and preparation for detailed characterizations were handled in the argon-filled glove box. The obtained $(TBA)_x FeSe$ single crystal keeps a well-defined shape with almost unchanged in-plane size and just the enlargement of c-axis compared with the pristine FeSe. The doping amount $x$ in $(TBA)_x FeSe$ determined by the electrochemical discharge curve is about 0.3, which is just a nominal value. The actual doping amount was determined using Electron Probe x-ray Micro-Analyzer (EPMA-8050G, Shimadzu), which is the most precise and accurate micro-analysis technique available. All elements from B to U can be analyzed with sensitivity at the level of ppm, which makes the determination of elemental composition from nm to mm thick layers possible. Measurements were made at several locations on each sample, and the averaged doping level was determined as $x \sim 0.15\pm0.03$ for $(TBA)_x FeSe$ and $x \sim 0.18\pm0.05$ for $(CTA)_x FeSe$. Other physical characterizations including the FTIR spectroscopy, X-ray diffraction and TEM have been employed to confirm the structure model of $(TBA)_x FeSe$ and $(CTA)_x FeSe$. More details about the crystal growth and characterization have been described in the Supplementary Information and a reported paper*(28,29)*.

**Electrical transport, magnetization and heat capacity measurements.** The electrical transport measurements were performed with a Physical Property Measurement System (PPMS-9T, Quantum Design) by a standard dc four-probe method. The sample was easily cleaved, so it was cut into long stripes and followed by attaching four silver wires with silver paint on the freshly cleaved surface. Considering the air-sensitivity of $(TBA)_x FeSe$, the sample was coated with a thin layer of chemically stable high-vacuum silicone grease before taken out of the glove box to be measured. The electrical resistance was measured upon both warming and cooling processes in order to ensure no temperature effect from the electrodes on the sample. The V (I) curves were measured on the PPMS with an electrical transport option (ETO). The voltages are in the range of tens of mV while scanning current within tens of mA.

Magnetization measurements were carried out with a commercial SQUID-VSM(7 T, Quantum Design). In order to reduce the magnetic background, a specified quartz paddle sample holder with extremely weak magnetic signal was used to hold the sample. The (TBA)$_x$FeSe was glued to the holder in a horizontal or vertical orientation with a little GE varnish and then coated with high-vacuum silicone grease. The magnetization was measured in FC and ZFC modes at magnetic field of 5 Oe and 7 T. The superconducting shielding fraction was estimated at 5 K from ZFC curve, which is almost 100% considering the geometrical demagnetization factor for H//c. The different magnitudes of χ under H//ab and H//c indicate a significant 2D characteristic in (TBA)$_x$FeSe.

The heat capacity measurements were performed in a 14 T Quantum Design Dynacool PPMS System. The thermal relaxation technique was employed with the magnetic field applied along c-axis.

**Nuclear magnetic resonance measurements.** Standard NMR spin-echo technique was applied with a commercial NMR spectrometer from Thamway Co. Ltd. The external magnetic field was generated by a 12 T magnet from Oxford Instruments. In order to greatly improve the measurement accuracy and efficiency, here a 50% $^{77}$Se isotope-enriched sample was used in the NMR measurements. (TBA)$_x$Fe$^{77}$Se single crystal was put into a prepared NMR coil made from copper and then the coil was sealed using epoxy glue so as to avoid the sample exposure to the air. The assembling process was performed in the glove-box filled with Ar gas. The $^{63}$Cu NMR signal from copper coil was used to calibrate the external field. The $^{77}$Se NMR spectra were obtained by fast Fourier transform (FFT) method. The nuclei of $^{77}$Se have a nuclear spin $I = 1/2$ and the Knight shift was extracted from the NMR central transition line using the gyromagnetic ratio $\gamma/2\pi = 8.118$ MHz/T.

The $^{77}$Se spin-lattice relaxation time (T$_1$) measurements were carried out by using the saturation-recovery method at the peak positions of NMR spectra at all temperatures, and the spin-echo decay was fitted by a simple exponential function $m(t) = A_0 + A\exp[-(t/T_1)^r]$. The curve is fitted very well and T$_1$ is of single component. Error bars are determined by least square fittings to the experimental data. 1/T$_1$ is related to the dynamical susceptibility and is expressed as $\frac{1}{T_1 T} = \frac{2\gamma_n^2 k_B}{(\gamma_e \hbar)^2} \sum_q |A_q|^2 \frac{\chi''_\pm(q,\omega_0)}{\omega_0}$, where $\gamma_{n\,(e)}$ is the gyromagnetic ratio of the nuclear spins (electronic spins), $A_q$ is the hyperfine coupling tensor, $\chi''_\pm(q,\omega_0)$ is the imaginary part of the dynamical susceptibility in the direction perpendicular to the quantization axis, and $\omega_0 = \gamma_n H$ is the resonance frequency under a magnetic field $\mu_0 H$. Because $\frac{1}{T_1 T} \propto (N(E_F))^2$ for conventional metals, the value of $1/T_1 T$ is a good quantity to study the superconducting gap.

**STM/STS measurements.** The STM experiment was conducted in a Createc LT-STM. The sample was cleaved in an ultra-high vacuum at liquid nitrogen temperature and immediately transferred into the STM part. All STM investigations reported here were acquired at 5 K. The tungsten tips were treated and calibrated using a Au(111) surface. The conductance spectra were taken by using a lock-in amplifier with a modulation voltage of 1 mV and at a frequency of 857 Hz.

**Nernst effect measurements.** The Nernst signal of 13.5 Tesla was measured by steady state measurement. The size of our sample is 2×0.6×0.2 mm$^3$. The cold end of the sample was glued to an alumina flake with silver glue and then the alumina flake was fixed on the heat-sunk copper block. A 1.8 kΩ heater, fixed on the hot end of sample, was used to generate heat flowing. The longitudinal thermal gradient was measured using two type-E thermocouples. The final Nernst data was obtained by the symmetric of the measured voltage with positive and negative magnetic fields: $e_y = E_y/(\partial T/\partial x) = [\Delta V_y(H)/\Delta T_x - \Delta V_y(-H)/\Delta T_x](L/2w)$, where $\Delta V$ is the difference in the voltage measured with and without thermal gradient. L is the length (along the x-axis) and w the width (along the y-axis) of the sample. This procedure removes any thermoelectric contribution from the sample or the rest of the measurement circuit.

# Figures and Captions

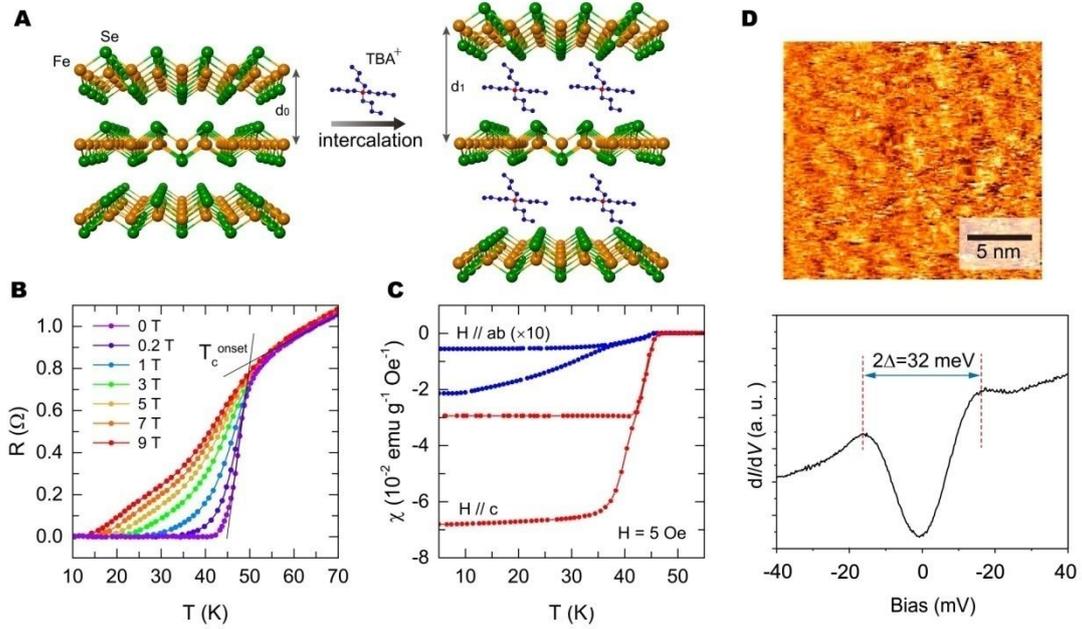

**Fig. 1. Crystal structure and superconducting properties of $(TBA)_xFeSe$. (A):** The crystal structure of the pristine FeSe with c-axis lattice constant $d_0 \sim 5.5$ Å and the intercalated $(TBA)_xFeSe$ with $d_1 \sim 15.5$ Å. **(B):** Temperature dependence of in-plane resistance under different magnetic fields applied along c-axis. The $T_c^{onset}$ is determined by the intersection of the linear extrapolation of normal-state resistance $R_n$ and the sharp superconducting transition, and the $T_{c0}$ is determined by using the 1% normal state resistance criterion. The fan-shaped broadening of resistive transition under magnetic fields indicates a strong 2D characteristic. **(C):** Temperature dependence of anisotropic magnetic susceptibility measured in FC and ZFC modes under magnetic field of 5 Oe applied along in-plane (blue) and out-of-plane (red), respectively. The significant difference of shielding fraction between two orientations, usually up to dozens of times, supports a strong 2D characteristic. **(D):(Upper panel)** Atomically flat STM image for $(TBA)_xFeSe$ with a bias voltage of $V_{bias}= 1$ V and tunneling current of $I_t= 220$ pA. **(Lower panel)** Tunneling spectrum taken on a cleaved surface of $(TBA)_xFeSe$ at 5 K reveals the appearance of a superconducting gap. Pronounced superconducting coherence peaks appear at ±16 meV.

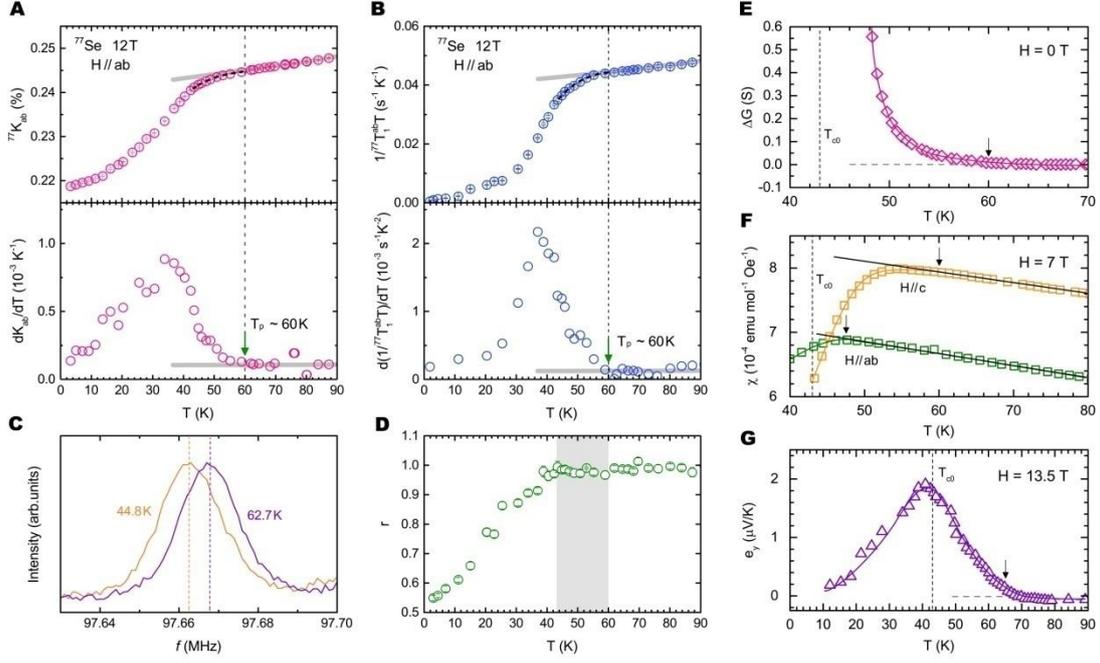

**Fig. 2. Evidence for preformed Cooper pairs above $T_{c0}$ in (TBA)$_x$FeSe. (A):** Temperature dependence of Knight shift $K$ (upper panel) and its first derivative (lower panel). **(B):** Temperature dependence of the spin-lattice relaxation rate divided by temperature $1/T_1T$ (upper panel) and its first derivative (lower panel). The external magnetic field of 12 Tesla in **(A)** and **(B)** was applied within the ab-plane. **(C):** The NMR spectra of $^{77}$Se nuclei measured at 44.8 K and 62.7 K, respectively. The uniform frequency shift of NMR spectrum to lower frequency below $T_p$ without obvious distortion or broadening indicates an intrinsic change of Knight shift below $T_p$. **(D):** Temperature evolution of stretching exponent $r$. The value of $r$ is extracted from the $T_1$ fitting by the stretched exponential function $m(t) = A_0 + A\exp[-(t/T_1)^r]$. The $r$ maintains a constant value of ~ 1 in the temperature range between $T_{c0}$ and $T_p$ (the shaded region). The deviation at lower temperature is ascribed to vortex in superconducting state. **(E):** Temperature dependence of conductance enhancement due to 2D superconducting fluctuations at zero field extracted from the data in Fig.1B. The conductance enhancement is defined as $\Delta G = 1/R(T) - 1/(R_n(T))$. $R_n$ is determined by the linear extrapolation of normal-state resistance at high temperature. The arrow shows the onset of the conductance enhancement at $T_p$. **(F):** The high-field magnetic susceptibility $\chi_{ab}$ and $\chi_c$ measured in FC mode with the magnetic field of 7 Tesla applied in-plane (green) and out-of-plane (orange), respectively. The black solid lines are the extrapolation fitting curves of high-temperature behavior. The arrow indicates the onset of diamagnetism. **(G):** Temperature dependence of Nernst effect under a magnetic field of 13.5 Tesla applied along c-axis. A vortex-related Nernst signal is observed well above $T_{c0}$. The arrow shows the onset of the vortex-related Nernst effect at ~ 65 K. It should be noted that the $T_p$ determined by Nernst effect is slightly higher than other probes. It suggests that the Nernst effect is more sensitive to detect superconducting fluctuation.

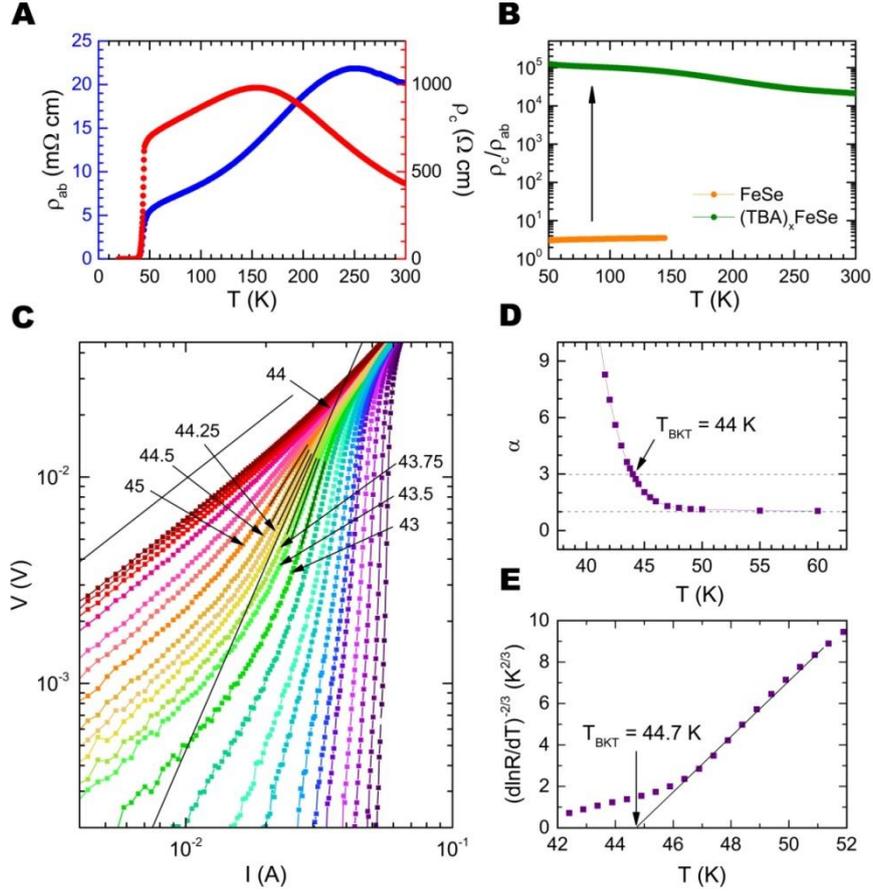

**Fig. 3. Two-dimensional electronic properties of (TBA)$_x$FeSe. (A):** The temperature dependence of in-plane and out-of-plane resistivity. **(B):** The anisotropy of resistivity ($\rho_c/\rho_{ab}$) for (TBA)$_x$FeSe and the pristine FeSe. The data of FeSe is adopted from the reference *(40)*. **(C):** $V(I)$ curves at various temperatures plotted in a logarithmic scale. The numbers provide the measured temperature for each curve. The short black lines are fits of the data across the $T_{c0}$. The two long black lines correspond to $V \sim I$ and $V \sim I^3$ behavior, respectively. The temperature for $V \sim I^3$ behavior is defined as $T_{BKT} \sim$ 44 K. **(D):** Temperature dependence of the power-law exponent $\alpha$, as deduced from the fits shown in **(C)**. **(E):** $R$-$T$ dependence for the (TBA)$_x$FeSe single crystal ($I$ = 500 μA), plotted in a $[d\ln(R)/dT]^{-2/3}$ scale. The solid line shows the fitting to the Halprin-Nelson formula $R(T) = R_0 exp[-b/(T - T_{BKT})^{1/2}]$ with $T_{BKT} \sim$ 44.7 K.

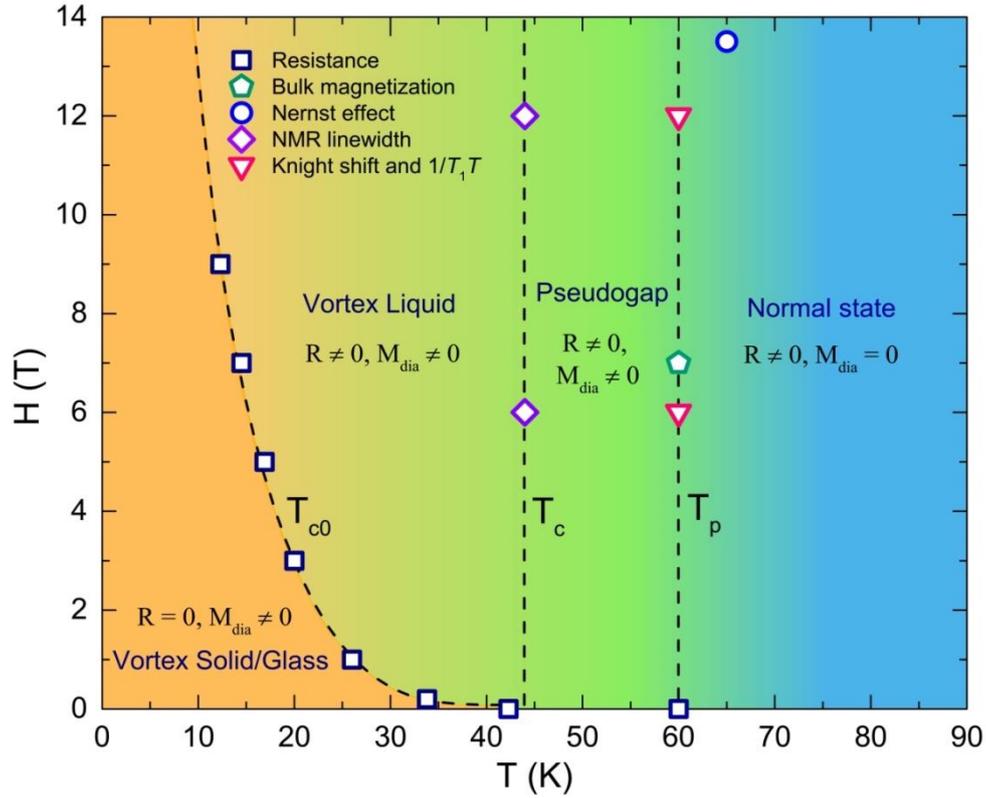

**Fig. 4. A summarized *H-T* phase diagram for (TBA)$_x$FeSe.** The external magnetic field is parallel to the c-axis. $T_{c0}$ stands for zero resistance temperature. $T_c$ stands for the superconducting temperature defined by the onset temperature of NMR line broadening. $T_p$ stands for the onset temperature of pseudogap behavior. Resistivity, magnetization and NMR measurements give the same $T_p$ of about 60 K. However, the Nernst effect gives a slightly higher $T_p$ of about 65 K. This suggests that Nernst effect is more sensitive than other probes to detect superconducting fluctuations. In the vortex solid/glass state, the resistivity becomes zero with a highly anisotropic diamagnetic signal. In the vortex liquid state, the resistivity is no longer zero but there is still anisotropic diamagnetic signal. Such phase diagram for vortex is similar to that in high-$T_c$ cuprate superconductors(*41*), in which the vortex physics is strongly affected by the 2D superconducting fluctuations (SCFs). Above $T_c$, a remarkable pseudogap phenomenon appears up to $T_p$.